\documentclass[a4paper,12pt]{article}

\usepackage{epsfig}
\usepackage{amstex}
\usepackage{amssymb}

\begin{document}


\renewcommand{\abstractname}{Abstract}
\renewcommand{\figurename}{Figure}

\providecommand{\nt}{\notag \\}

\providecommand{\SP}{\scriptstyle}

\providecommand{\stl}{\tilde{t}_L}
\providecommand{\str}{\tilde{t}_R}
\providecommand{\ste}{\tilde{t}_1}
\providecommand{\stz}{\tilde{t}_2}
\providecommand{\sti}{\tilde{t}_i}
\providecommand{\stj}{\tilde{t}_j}
\providecommand{\stez}{\tilde{t}_{1,2}}
\providecommand{\st}{\tilde{q}}
\providecommand{\gt}{\tilde{g}}

\providecommand{\tb}{\bar{t}}
\providecommand{\steb}{\bar{\tilde{t}}_1}
\providecommand{\stzb}{\bar{\tilde{t}}_2}
\providecommand{\stib}{\bar{\tilde{t}}_i}
\providecommand{\stjb}{\bar{\tilde{t}}_j}

\providecommand{\sten}{\tilde{t}_{1 0}}
\providecommand{\stzn}{\tilde{t}_{2 0}}
\providecommand{\stin}{\tilde{t}_{i 0}}
\providecommand{\stjn}{\tilde{t}_{j 0}}
\providecommand{\chijn}{\tilde{\chi}_j^0}
\providecommand{\chijp}{\tilde{\chi}_j^+}
\providecommand{\chien}{\tilde{\chi}_1^0}
\providecommand{\chizn}{\tilde{\chi}_2^0}
\providecommand{\chiep}{\tilde{\chi}_1^+}

\providecommand{\mse}{m_{\tilde{t}_{\SP 1}}}
\providecommand{\msz}{m_{\tilde{t}_{\SP 2}}}
\providecommand{\msi}{m_{\tilde{t}_{\SP i}}}
\providecommand{\msj}{m_{\tilde{t}_{\SP j}}}
\providecommand{\msez}{m_{\tilde{t}_{\SP 1,2}}}
\providecommand{\msen}{m_{\tilde{t}_{\SP 1 0}}}
\providecommand{\mszn}{m_{\tilde{t}_{\SP 2 0}}}
\providecommand{\msjn}{m_{\tilde{t}_{\SP j 0}}}
\providecommand{\mg}{m_{\tilde{g}}}
\providecommand{\mt}{m_t}
\providecommand{\ms}{m_{\tilde{q}}}
\providecommand{\ml}{\lambda}

\providecommand{\mgq}{m_{\tilde{g}}^2}
\providecommand{\mtq}{m_t^2}

\providecommand{\pg}{p_{\tilde{g}}}
\providecommand{\pt}{p_t}
\providecommand{\ps}{p_{\tilde{t}_{\SP 1}}}

\providecommand{\tmixn}{\tilde{\theta}_0}
\providecommand{\tmix}{\tilde{\theta}}
\providecommand{\delZ}{\delta \tmix}

\providecommand{\ZM}{Z^{1/2}}
\providecommand{\MM}{{\cal M}^2}

\providecommand{\as}{\alpha_s}
\providecommand{\sw}{\sin^2 \theta_W} 
\providecommand{\msbar}{\overline{{\rm MS}}}
\providecommand{\ee}{\epsilon}
\providecommand{\fac}{{\cal N}}

\providecommand{\szt}{s_{2\tmix}}
\providecommand{\czt}{c_{2\tmix}}
\providecommand{\svt}{s_{4\tmix}}
\providecommand{\sztsq}{s_{2\tmix}^2}
\providecommand{\cztsq}{c_{2\tmix}^2}

\providecommand{\mat}{|{\cal M}|^2}
\providecommand{\MB}{\; |{\cal M}_B|^2 \;}
\providecommand{\MBa}{\; |{\cal M}_B|^2_{ang} \;}

\providecommand{\Bp}{\dot{B}}

\providecommand{\sigzt}{\sigma_{2\tmix}}

\providecommand{\Ce}{C(\ste,t,t,\gt,\ste)}
\providecommand{\Cz}{C(\ste,t,t,\gt,\stz)}

\providecommand{\mug}{\mu_{\tilde{g}}}
\providecommand{\mut}{\mu_t}
\providecommand{\mue}{\mu_1}
\providecommand{\muz}{\mu_2}
\providecommand{\muez}{\mu_{1 2}}
\providecommand{\mugs}{\mu_{\tilde{g} \tilde{q}}}
\providecommand{\muge}{\mu_{\tilde{g} 1}}
\providecommand{\mugz}{\mu_{\tilde{g} 2}}
\providecommand{\muet}{\mu_{1 t}}
\providecommand{\muzt}{\mu_{2 t}}
\providecommand{\mugte}{\mu_{\tilde{g} t 1}}
\providecommand{\mugtz}{\mu_{\tilde{g} t 2}}
\providecommand{\muteg}{\mu_{t 1 \tilde{g}}}
\providecommand{\mutzg}{\mu_{t 2 \tilde{g}}}
\providecommand{\muget}{\mu_{\tilde{g} 1 t}}
\providecommand{\muegt}{\mu_{1 \tilde{g} t}}
\providecommand{\muzgt}{\mu_{2 \tilde{g} t}}
\providecommand{\mutez}{\mu_{t 1 2}}
\providecommand{\real}{{\cal\text{Re\,}}}
\providecommand{\psm}{\ \raisebox{0.8mm}{\scriptsize $+$}\!/\!
                      \raisebox{-0.5mm}{\scriptsize $-$}\ }

\renewcommand{\thefootnote}{\fnsymbol{footnote}}

\begin{titlepage}

\begin{flushright}
DESY 96--178 \\
October 1996 \\
\end{flushright}

\vspace{1cm}

\begin{center}
\baselineskip25pt

\def\thefootnote{\fnsymbol{footnote}}
{\large\sc Stop decays in susy-qcd\footnote{partially supported by EU 
                                            contract CHRX--CT--92--0004}  }

\end{center}

\setcounter{footnote}{3}

\vspace{1cm}

\begin{center}
\baselineskip12pt

{\sc 
W.~Beenakker$^1$\footnote{Fellow of the Royal Dutch Academy 
                          of Arts and Sciences},
R.~H\"opker$^2$,
T.~Plehn$^2$
and P.~M.~Zerwas$^2$} \\ 
\vspace{1cm}

$^1$ Instituut--Lorentz, P.O.~Box 9506, 
     NL--2300 RA Leiden, The Netherlands \\

\vspace{0.3cm}

$^2$ Deutsches Elektronen--Synchrotron DESY, 
     D--22603 Hamburg, FRG 

\end{center}

\vspace{2cm}
\begin{abstract}
  \normalsize \noindent The partial widths are determined for stop
  decays to top quarks and gluinos, and gluino decays to stop
  particles and top quarks (depending on the masses of the particles
  involved). The widths are calculated including one-loop SUSY-QCD
  corrections. The radiative corrections for these strong-interaction
  decays are compared with the SUSY-QCD corrections for electroweak
  stop decays to quarks and neutralinos/charginos and top-quark decays
  to stops and neutralinos.
\end{abstract}

\end{titlepage}

\def\thefootnote{\arabic{footnote}}
\setcounter{footnote}{0}

\setcounter{page}{2}

\section{Introduction} 

The top and stop particles form a complex system in supersymmetric
theories. The strong Yukawa coupling between top/stop and
Higgs fields gives rise to a large mixing of the $L$ and $R$ stop
states $\stl $ and $\str $, which are associated with the left and
right chiral top-quark states.  The mass splitting between the stop mass 
eigenstates $\ste $ and $\stz $ can therefore be quite large. In fact,
it is possible that the mass $\mse $ of the lightest stop
state is even smaller than the top mass $m_t $ itself \cite{stop}.

Depending on the mass values of the particles involved, quite different
decay scenarios will be realized in the stop--top sector. If stop
particles are very heavy, they can decay into top quarks and gluinos,
\begin{equation} 
\stj \rightarrow t + \gt \qquad \qquad \qquad \left[ \msj > \mt + \mg \right] 
\label{eq_st1} 
\end{equation} 
In this paper we generalize the analysis of Ref.\cite{roland}, in
which the decay widths for the squarks related to the light quark
species ($\st = \tilde{u},...,\tilde{b}$) were calculated, to the
stop-decay processes (\ref{eq_st1}) in next-to-leading order SUSY-QCD;
in this more complex case the stop mixing and the non-zero top-quark
mass in the final state must be taken into account. In a similar way
we analyze the crossed channel
\begin{equation} 
\gt \rightarrow \tb + \stj \quad \text{and c.c.} \qquad 
    \left[ \mg > \mt + \msj \right] 
\label{eq_gt} 
\end{equation} 
in leading and next-to-leading order. For small stop masses, in
particular for $\ste $, the decay channel (\ref{eq_st1}) is presumably
shut kinematically and decays to quarks and light neutralinos or
charginos ($\ste \rightarrow t\chien, b\chiep $) will be dominant
\cite{vienna}.\footnote{For higher-order electroweak stop decays we
  refer to the recent paper Ref.\cite{vienna2}.} For the sake of
comparison with Refs.\cite{vienna,karlsruhe}, we re-analyze also these
decay modes in next-to-leading order SUSY-QCD.  Finally, in the
exceptional case $\mt > \mse $, the interesting decay mode $t
\rightarrow \ste \chien $ may occur, see e.g. Ref.\cite{borzumati};
the partial width of this non-standard top decay has recently been
determined in next-to-leading order SUSY-QCD in Ref.\cite{top_decay}.

\section{Theoretical Set-up}

To lowest order the partial widths for the stop and gluino decays
(\ref{eq_st1}) and (\ref{eq_gt}) are given by\footnote{ As usual,
  $\kappa = ( \sum_i m_i^4 - \sum_{i \neq j} m_i^2 m_j^2 )^{1/2} $,
  the sums running over all particles involved in the decay process.}

\vspace{-0.2cm}

\begin{alignat}{2}
\Gamma (\stez \rightarrow t\,\gt) &= \frac{2 \as \kappa}{ 3 \msez^3} 
  \left[ \msez^2 - m_t^2 - \mg^2 \pm 2 m_t \mg \sin(2 \tmix) \right]  
\label{eq_gamlo_stj} \\
\Gamma (\gt \rightarrow \tb\,\stez) &={}- \frac{\as \kappa}{ 8 \mg^3} 
  \left[ \msez^2 - m_t^2 - \mg^2 \pm 2 m_t \mg \sin(2 \tmix) \right] 
\label{eq_gamlo_gt}  
\end{alignat}
Here $\mse $ and $\msz $ are the eigenvalues of the stop mass
matrix\footnote{ The sign conventions follow the SPYTHIA program
  \cite{spythia}, which is based on Ref.\cite{drees}.}
\cite{stop,haber}:
\begin{eqnarray}
\MM &=& 
\left(  \begin{array}{cc} 
 \MM_{LL} & \MM_{LR} \\ \MM_{RL} & \MM_{RR}
        \end{array}  \right)  \notag \\[0.2cm]
&=&
\left(  \begin{array}{cc} 
 m_Q^2+m_t^2+
   \left( \frac{1}{2} - \frac{2}{3} s_w^2 \right) m_Z^2 \cos(2 \beta) & 
 -m_t \left( A_t + \mu \cot \beta \right) \\ 
 -m_t \left( A_t + \mu \cot \beta \right)  &  
 m_U^2+m_t^2+ \frac{2}{3} s_w^2 m_Z^2 \cos(2 \beta) 
        \end{array}  \right)  
\label{eq_massmatrix}
\end{eqnarray}
The quantities $m_Q,m_U,\mu$, and $A_t$ are the usual soft
SUSY-breaking mass and trilinear parameters, $m_Z$ and $s_w$ are the
$Z$-boson mass and the weak mixing angle, and $\tan\beta$ is the ratio
of the two vacuum expectation values in the Higgs sector.  The
diagonal entries of the stop mass matrix correspond to the $L$ and $R$
squark-mass terms, the off-diagonal entries are due to chirality-flip
Yukawa interactions.  The chiral states $\stl $ and $\str $ are
rotated into the mass eigenstates $\sten$ and $\stzn$
\begin{eqnarray}
\left(  \begin{array}{c} 
 \sten \\ \stzn 
        \end{array}  \right)
= \left(  \begin{array}{cc} 
 \cos \tmixn & \sin \tmixn \\
 - \sin \tmixn & \cos \tmixn
        \end{array}  \right)  
\left(  \begin{array}{c} 
 \stl \\ \str 
        \end{array}  \right)
\end{eqnarray}
by these Yukawa interactions.
The mass eigenvalues and  the rotation angle
can be calculated from the mass matrix (\ref{eq_massmatrix}):
\begin{alignat}{2}
&m^2_{\tilde{t}_{\SP 1 0}},m^2_{\tilde{t}_{\SP 2 0}} =
           \frac{1}{2} \left[ 
           \MM_{LL}+\MM_{RR}
            \mp \left[ (\MM_{LL}-\MM_{RR})^2 + 4 (\MM_{LR})^2 \right]^{1/2} 
                              \right] \label{mstopLO}\\
&\sin(2 \tmixn) = \frac{2 \MM_{LR}}{\msen^2 - \mszn^2} \quad \text{and} \quad
 \cos(2 \tmixn) = \frac{\MM_{LL}-\MM_{RR}}{\msen^2 - \mszn^2} \label{thetaLO}
\end{alignat}
By definition we take $\sten$ to correspond to the lightest stop state.

The mixing angle is an observable quantity, as evident from the decay
widths (\ref{eq_gamlo_stj}) and (\ref{eq_gamlo_gt}). Since the widths of
supersymmetric particles are notoriously difficult to measure,
production processes may instead be adopted for the operational
definition of the mixing angle in practice. Pair production in $e^+
e^-$ collisions, $e^+ e^- \rightarrow \sti \stjb $, lends itself as a
convenient observable \cite{leptoprod}.

\begin{figure}[h]
\epsfig{file=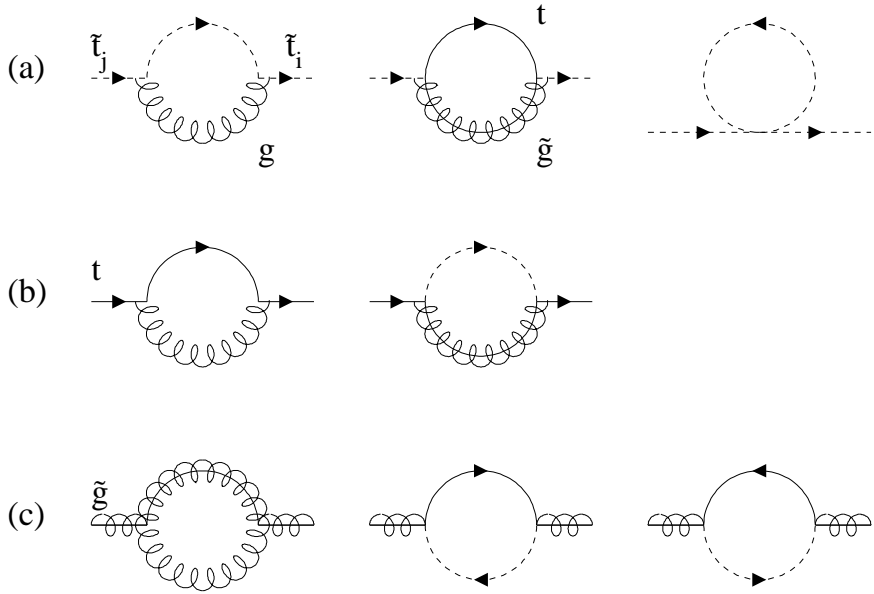,width=12cm} 
\caption[]{The Feynman diagrams for the self-energies:
(a) self-energy of the stop particles, including the mixing due to the 
    second and third diagram; 
(b) top-quark self-energy; 
(c) gluino self-energy [including fermion-number violation].}
\label{fig_self}
\end{figure}

SUSY-QCD corrections, as exemplified by the diagrams of
Fig.\ref{fig_self}(a), modify the stop mass matrix and the fields,
necessitating the renormalization of the masses $ \msjn^2 = \msj^2 +
\delta \msj^2 $ and of the wave functions $ \stin = \ZM_{ij} \stj $;
the renormalization of the mixing angle can be related to the
renormalization matrix $\ZM$. Without loss of generality, we may
assume $\ZM_{12}= -\ZM_{21}$ since, as will be shown later, the
reduced self-energy matrix $\Sigma_{ij}/(\msi^2-\msj^2)$ is
antisymmetric.  In that case the renormalization matrix $\ZM $ can be
written to order $\as$ as the product of a diagonal matrix with the
elements $\ZM_{jj} = 1+\delta Z_{jj}/2 $ and a rotation matrix
parametrized by the small angle $\delZ$:
\begin{eqnarray}
\ZM \mapsto 
\left(  \begin{array}{cc} 
    1 + \delta Z_{11}/2 & 0 
 \\ 0 & 1 + \delta Z_{22}/2
        \end{array}  \right)  
\left(  \begin{array}{cc} 
     \cos \delZ & \sin \delZ 
 \\ -\sin \delZ & \cos \delZ
        \end{array}  \right)  
\end{eqnarray}
with $\delZ =\ZM_{12}= -\ZM_{21}$.  Within this formalism the rotational
part of $\ZM $ can be interpreted as a shift in the mixing angle,
$\tmixn \mapsto \tmixn - \delZ \equiv \tmix$. The remaining renormalization 
constants are fixed by imposing the following two conditions: 
\medskip \\ 
(i) The real part of the diagonal elements in the stop propagator matrix
$\Delta_{ij}(p^2)$ develops poles for $p^2 \rightarrow \msj^2$, with
the residues being unity. This requirement fixes the counterterms
$\delta \msj^2 $ and the diagonal wave-function renormalization
$\delta Z_{jj}$. The so-defined renormalized stop masses $\msj^2$ are
called pole masses.  
\medskip \\ 
(ii) We define a running mixing angle
$\tmix(Q^2)$ by requiring that the real part of the off-diagonal
elements of the propagator matrix $\Delta_{ij}(p^2)$ vanishes for a
specific value $p^2=Q^2$ of the four-momentum squared. In this scheme
the [real or virtual] particles $\ste $ and $\stz $ propagate
independently of each other at four-momentum squared $Q^2$ and do not
oscillate.

The dependence of the renormalized mixing angle $\tmix$ on the
renormalization scale $Q$ is indicated by the notation $\tmix(Q^2)$.
Different choices for $Q^2$ are connected by a finite shift in
$\tmix(Q^2)$, which is calculated in the next paragraph.  Quite often
the renormalized mixing angle is defined in an 'on-shell
renormalization scheme' \cite{karlsruhe,vienna_prod} in which the
renormalization of the mixing angle cannot be linked to the stop wave
functions any more. Even though the definitions of the mixing angle
are different in the two schemes, the physical observables, widths and
cross-sections are of course equivalent to ${\cal O}(\as)$.

\begin{figure}[h]
\epsfig{file=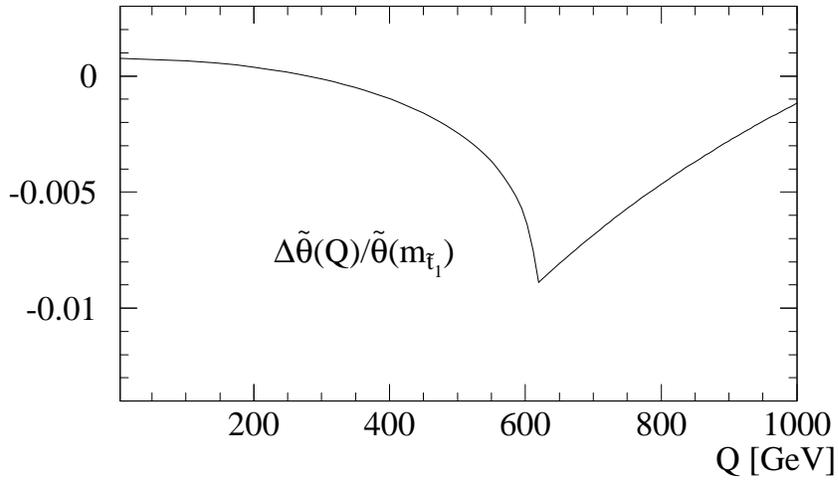,width=12cm} 
\caption[]{
The dependence of $\tmix(Q^2)$ on the renormalization scale $Q$. The 
normalized shift is shown relative to $\tmix(\mse^2)$: 
$[\tmix(Q^2)-\tmix(\mse^2)]/\tmix(\mse^2)$.
The input mass values are the same as for the stop 
decay to gluinos: $m_{1/2}=150$~GeV, $m_0=800$~GeV, $A_0=200$~GeV, $\mu>0$, 
for which the leading-order mixing angle is given by $1.24$ rad.
The minimum of the correction corresponds to the threshold $Q=\mg+\mt$ 
in the scalar integral.}
\label{fig_ang}
\end{figure}

In carrying out this renormalization program we find the following
expressions for the various counter terms: 
\begin{displaymath}
\delta \msj^2 = \real\Sigma_{jj}(\msj^2) \qquad
\delta Z_{jj} = -\real\dot{\Sigma}_{jj}(\msj^2) \qquad
\delta \tmix = - \real\Sigma_{12}(Q^2)/[\msz^2-\mse^2]  
\end{displaymath}
Here $\Sigma_{ij}(p^2)$ and $\dot{\Sigma}_{ij}(p^2) =
\partial \Sigma_{ij}(p^2)/ \partial p^2 $ denote the unrenormalized 
self-energy matrix and its derivative [see also
Ref.~\cite{vienna_prod,karlsruhe}] :
\begin{alignat}{2}
  \Sigma_{12}(p^2) &= - 2 \pi C_F \as \Big\{ \svt \left[ A(\msz) -
      A(\mse) \right] + 8 \mg m_t \czt \,B(p^2,\mg,\mt) \Big\} \\[1mm]
  \Sigma_{21}(p^2) &= \Sigma_{12}(p^2) \\[1mm] 
  \Sigma_{11}(p^2) &= - 4 \pi
      C_F \as \Big\{ \left(1+\czt^2 \right) A(\mse) + \szt^2 \,A(\msz)
      \notag \\[1mm] & \phantom{= - 4 \pi i C_F \as } 
      {}-2\,A(\mg) - 2\,A(\mt) 
      -2 \left( p^2 + \mse^2 \right) B(p^2,\lambda,\mse) \notag \\[1mm]
      &\phantom{= - 4 \pi i C_F \as } {}+2 \left( p^2 - \mg^2 - \mt^2 
      + 2 \mg \mt \szt \right) B(p^2,\mg,\mt) \Big\} \\[1mm] 
  \Sigma_{22}(p^2) &= -
      4 \pi C_F \as \Big\{ \left(1+\czt^2 \right) A(\msz) + \szt^2
      \,A(\mse) \notag \\[1mm] & \phantom{= - 4 \pi i C_F \as } 
      {}-2\,A(\mg) - 2\,A(\mt) 
      -2 \left( p^2 + \msz^2 \right) B(p^2,\lambda,\msz) \notag \\[1mm] 
      & \phantom{= - 4 \pi i C_F \as } {}+2 \left( p^2 - \mg^2 
      - \mt^2 - 2 \mg \mt \szt \right) B(p^2,\mg,\mt) \Big\}
\end{alignat}
[We have used the standard notation $s_\theta\equiv\sin \theta$ etc.]
The first two terms in $\Sigma_{12}(p^2)$, 
involving the $p^2$-independent 1-point function $A$, follow from
the third Feynman diagram in Fig.\ref{fig_self}(a), the remaining
term given by the 2-point function $B$, corresponds to the second
diagram in Fig.\ref{fig_self}(a). As both $A$ and $B$ are ultraviolet (UV)
divergent%
\footnote{The definitions of the scalar functions $A$ and $B$ can be 
          found in the Appendix.}, 
also $\delta \tmix$ is UV divergent; in $n=4-2\ee$ dimensions:
\begin{alignat}{2}
\delta \tmix \; |_{div} 
   =\:& \frac{C_F \as }
            {8 \pi (\mse^2 - \msz^2)}
    \left[ \svt \left(\msz^2 - \mse^2 \right) 
  + 8 \mg m_t \czt \right]\,\frac{1}{\ee}
\end{alignat} 
The change of the renormalized angle $\tmix(Q^2)$ between two
different values of $Q^2$ is finite:
\begin{equation}
  \tmix({Q'}^2)-\tmix(Q^2) 
  = {}- \frac{16 \pi C_F \as \mg \mt \cos(2 \tmix)}{\msz^2 - \mse^2}
    \,\real\!\left[ B({Q'}^2,\mg,\mt) - B(Q^2,\mg,\mt) \right] 
\label{eq_angle}
\end{equation}
This shift is independent of the regularization scheme. For
illustration the normalized shift relative to $\tmix(\mse^2)$:
$[\tmix(Q^2) - \tmix(\mse^2)]/\tmix(\mse^2)$, is shown in
Fig.\ref{fig_ang}.

\section{Stop and Gluino Decays}

\begin{figure}[h]
\epsfig{file=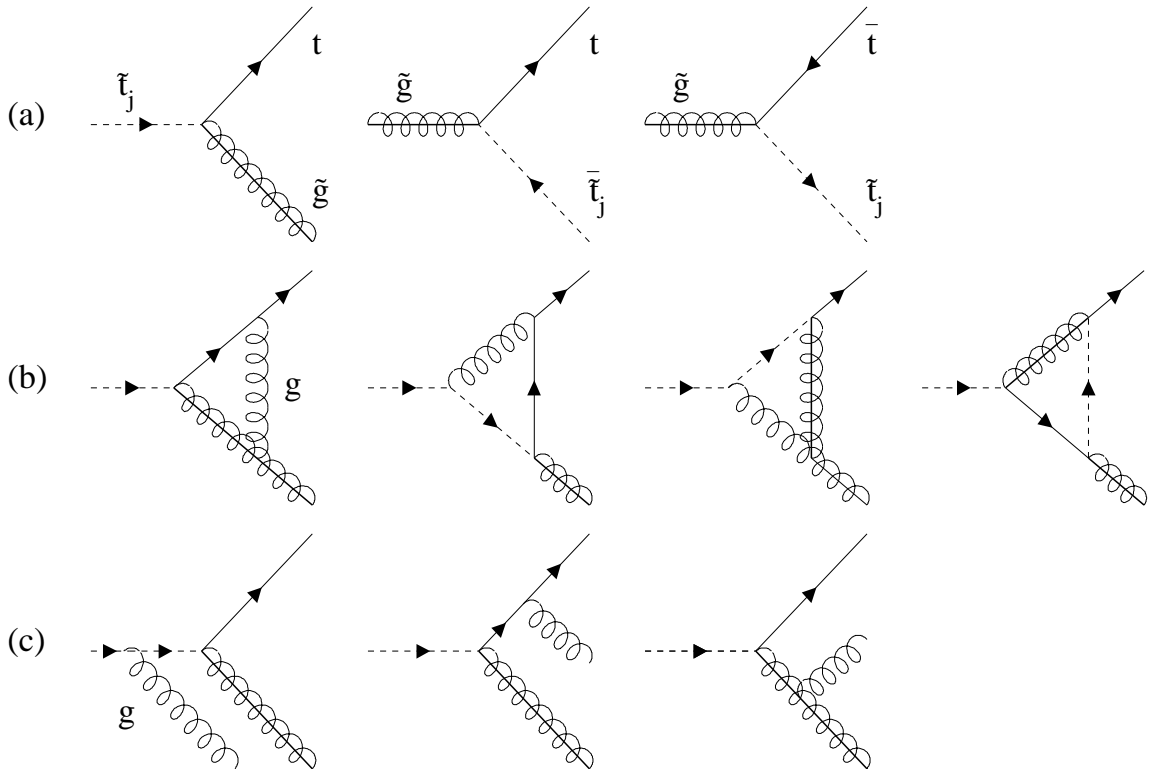,width=16cm} 
\caption[]{
(a) Born diagrams for stop and gluino decays; 
(b) vertex corrections for stop decays; 
(c) real-gluon emission for stop decays.
The corrections to gluino decays can be obtained by rotating the diagrams.} 
\label{fig_corr}
\end{figure}

The diagrams relevant for stop and gluino decays are presented in
Fig.\ref{fig_corr}. This set is complemented by the self-energy
diagrams for the stop particles, the gluinos, and the top quarks,
displayed in Fig.\ref{fig_self}. The Born diagrams are presented in
Fig.\ref{fig_corr}(a) for the two decay channels, the vertex
corrections in Fig.\ref{fig_corr}(b), and the hard-gluon radiation in
Fig.\ref{fig_corr}(c). In Figs.\ref{fig_corr}(b) and (c) only the
stop-decay diagrams are depicted, since gluino- and stop-decay
diagrams are related by crossing. The ultraviolet divergences are
regularized in $n$ dimensions, infrared and collinear
divergences\footnote{ Since no $ggg$ three-gluon vertices are
  involved in the calculation, infrared singularities can be
  regularized by a non-zero gluon mass.  This method can also be
  applied to SUSY-QCD processes including $g\gt\gt$ vertices of
  massive gluinos.} by introducing a small gluon mass $\lambda$.  The
renormalization of the strong coupling constants $g_s$ and $\hat{g}_s$
are carried out in the $\msbar$ renormalization scheme at the 
charge-renormalization scale $\mu_R$; a finite shift
\cite{shift} between the bare Yukawa coupling $\hat{g}_s$ and the bare
gauge coupling $g_s$ restores supersymmetry at the one-loop level in
the $\msbar$ scheme [see Ref.\cite{roland} for explicit verification]:
\begin{equation}
\hat{g}_s = g_s \left[ 1 + 
                       \frac{\as}{8\pi} \left( \frac{4}{3}N_c - C_F \right)
                \right]
\end{equation}
[Similarly the electroweak $\sti t \chijn$ and $\sti b \chijp$ couplings may 
be written as $a\,\hat{e} + b\,\hat{Y}_t + c\,\hat{Y}_b$, with 
$\hat{e} = e\,[1-\as C_F/(8\pi)]$ and $\hat{Y}_q = Y_q\,[1-3 \as C_F/(8\pi)]$
in terms of the electromagnetic coupling $e$ and the quark--Higgs Yukawa 
coupling $Y_q \propto e\,m_q$.]
The heavy particles (top quarks, squarks, gluinos) are 
removed from the $\mu_R^2$ evolution of $\as(\mu_R^2)$,
decoupled smoothly for momenta smaller than their masses. 
The masses of these heavy particles are defined as pole masses.
In the numerical analyses we have inserted the mass of the decaying
particle for the renormalization scale $\mu_R$.

The detailed analytic results for stop and gluino decays are presented
in the Appendix. In this section we illustrate the characteristic
features of the results by numerical evaluation of a few typical
examples, which properly reflect the size of the SUSY-QCD effects in
general. 

The masses and mixing parameters chosen in the examples are calculated
from the universal SUGRA parameters \cite{gut}: the common scalar mass
$m_0$, the common gaugino mass $m_{1/2}$, the trilinear coupling
$A_0$, the ratio of the vacuum expectation values of the Higgs fields
$\tan \beta$, and the sign of the higgsino mass parameter $\mu$. The
top-quark mass is set to $m_t=175$~GeV, and the Higgs parameter $\tan
\beta$ is fixed to 1.75. 
From this set, the pole masses of the
charginos, neutralinos, gluinos, and squarks, as well as the squark
mixing matrices can be calculated. We use the approximate formulae
implemented in SPYTHIA \cite{spythia}. We define a degenerate squark
mass by averaging over the five non--top flavors.  The renormalized
stop mixing angle $\tmix(\mse^2)$ is defined in this basis by imposing
the lowest-order relation Eq.\,(\ref{thetaLO}) in
terms of the renormalized (pole) masses.%
\footnote{Equivalently the mass-matrix parameters
  $m_Q$, $m_U$, and $A_t$ may be chosen as basic parameters, and 
  $\tmix(\mse^2)$ and $\msj$ may subsequently be
  defined in this basis by imposing the lowest-order relations 
  Eqs.\,(\ref{mstopLO}) and (\ref{thetaLO}).  
  We have checked that the two schemes are equivalent.} 
The mixing angles at other renormalization scales can be obtained by adding 
the appropriate finite shifts [see Eq.(\ref{eq_angle})]. 
\bigskip

\setlength{\parindent}{0pt}

{\boldmath $\text{\bf (a)} \; \stz \rightarrow t + \gt$}

\hspace{0.6cm} for $m_0=800$~GeV, $A_0=200$~GeV, $\mu>0$

\begin{figure}[h]
\epsfig{file=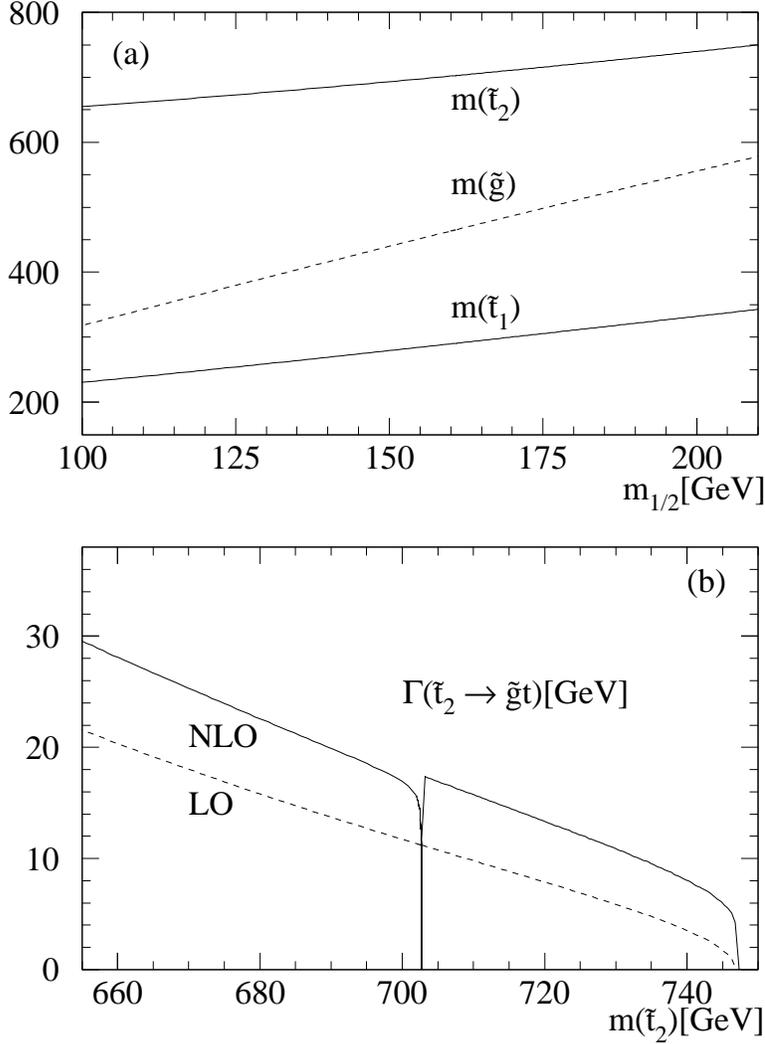,width=12cm} 
\caption[]{The SUSY-QCD corrections to the decay of heavy stop particles 
into top quarks and gluinos: 
(a) the masses of the particles [in GeV] as a function of the 
common gaugino mass $m_{1/2}$;
(b) the decay widths in leading order (dashed curve) and next-to-leading 
order (solid curve).
Input parameter set: $m_0=800$~GeV, $A_0=200$~GeV, $\mu>0$.
[The kink at the threshold for $\gt \to \bar{t}\ste$ can be smoothed out by
inserting the finite widths of the particles.]}
\label{fig_sg}
\end{figure}

In Fig.\ref{fig_sg}(a) the stop and gluino masses are given 
as a function of the common gaugino mass $m_{1/2}$. For the indicated set of 
parameters the decay $\stz \to t \gt$ is the only strong-interaction decay
mode that is kinematically allowed.

In Fig.\ref{fig_sg}(b) the width of $\stz$ is presented in lowest
order and in next-to-leading order as a function of $\msz$. Since in
this example $\mg$ rises faster than $\msz$ with increasing $m_{1/2}$,
the width drops to zero when $\msz$ is increased. The radiative
SUSY-QCD corrections vary between $+35 \%$, at the lower end of the
spectrum, and $\sim +100 \%$ at the upper end of the spectrum,
\emph{i.e.}, the corrections are large and positive. \bigskip

{\boldmath $\text{\bf (b)} \; \gt \rightarrow \ste \tb + \steb t$}

\hspace{0.6cm} for $m_0=400$~GeV, $A_0=200$~GeV, $\mu>0$

\begin{figure}[h]
\epsfig{file=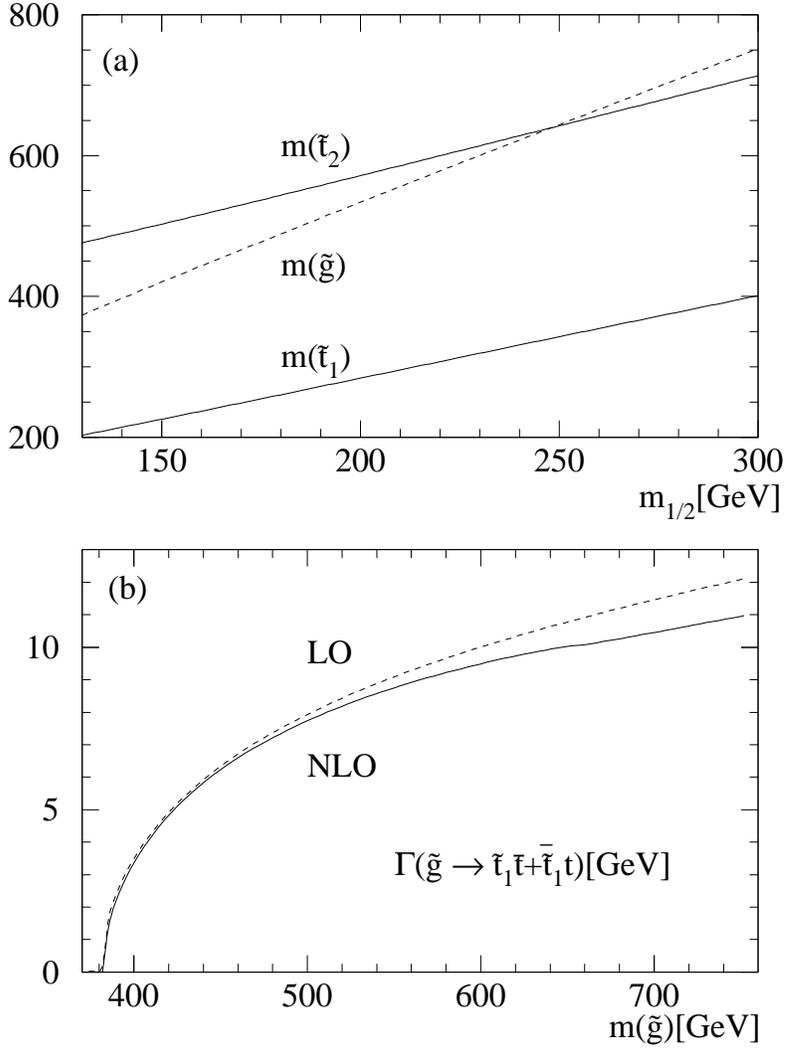,width=12cm} 
\caption[]{The SUSY-QCD corrections to the decay of gluinos into light 
stop particles and top quarks: 
(a) the masses of the particles [in GeV] as a function of the 
common gaugino mass $m_{1/2}$;
(b) the decay widths in leading order (dashed curve) and next-to-leading 
order (solid curve).
Input parameter set: $m_0=400$~GeV, $A_0=200$~GeV, $\mu>0$.}
\label{fig_gs}
\end{figure}

In a form analogous to the preceding example, masses and widths are
displayed in Figs.\ref{fig_gs}(a) and (b). Since $\mg$ rises faster
than $\mse$, the width increases with increasing gluino mass.
However, in contrast to stop decays, the SUSY-QCD corrections to
gluino decays are only modest and negative [$\sim - 10 \%$]. This
result is familiar from the analysis in Ref.\cite{roland}, where it
had been demonstrated analytically that negative $\pi^2$ terms,
arising from the crossing of diagrams, give rise to destructive
interference effects such that the overall correction is
small.\bigskip 

{\boldmath $\text{\bf (c)} \; \ste \rightarrow  t + \chijn \quad [j=1,...] 
                  \quad \text{\bf and} \quad 
                              \ste \rightarrow b + \chijp \quad[j=1,2] $}

\hspace{0.6cm} for $m_0=50$~GeV, $A_0=100$~GeV, $\mu<0$

\begin{figure}[h]
\epsfig{file=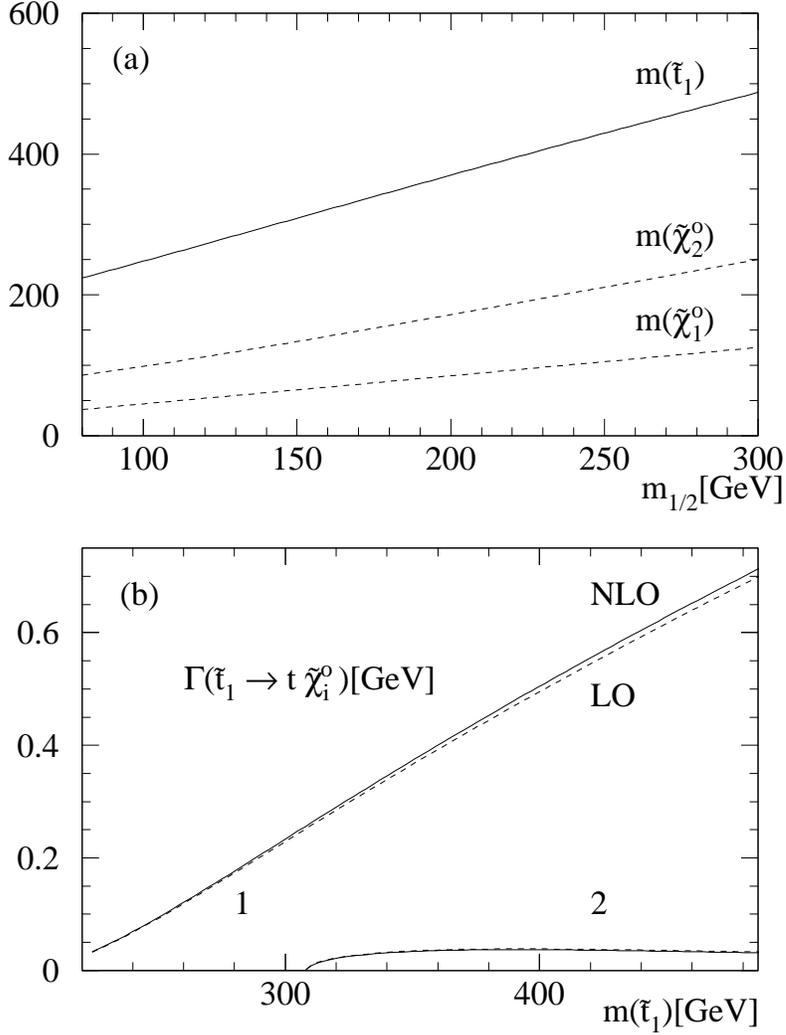,width=12cm} 
\caption[]{The SUSY-QCD corrections to the decay of light stop particles into
  top quarks and the possible neutralino eigenstates [see also
  Ref.\cite{vienna}]: (a) the masses of the particles [in GeV] as a
  function of the common gaugino mass $m_{1/2}$; (b) the decay widths
  in leading order (dashed curve) and next-to-leading order (solid
  curve).  Input parameter set: $m_0=50$~GeV, $A_0=100$~GeV, $\mu<0$.
  Only the decays into the two lightest neutralinos are kinematically
  allowed for this parameter set.}
\label{fig_sn}
\end{figure}

\begin{figure}[h]
\epsfig{file=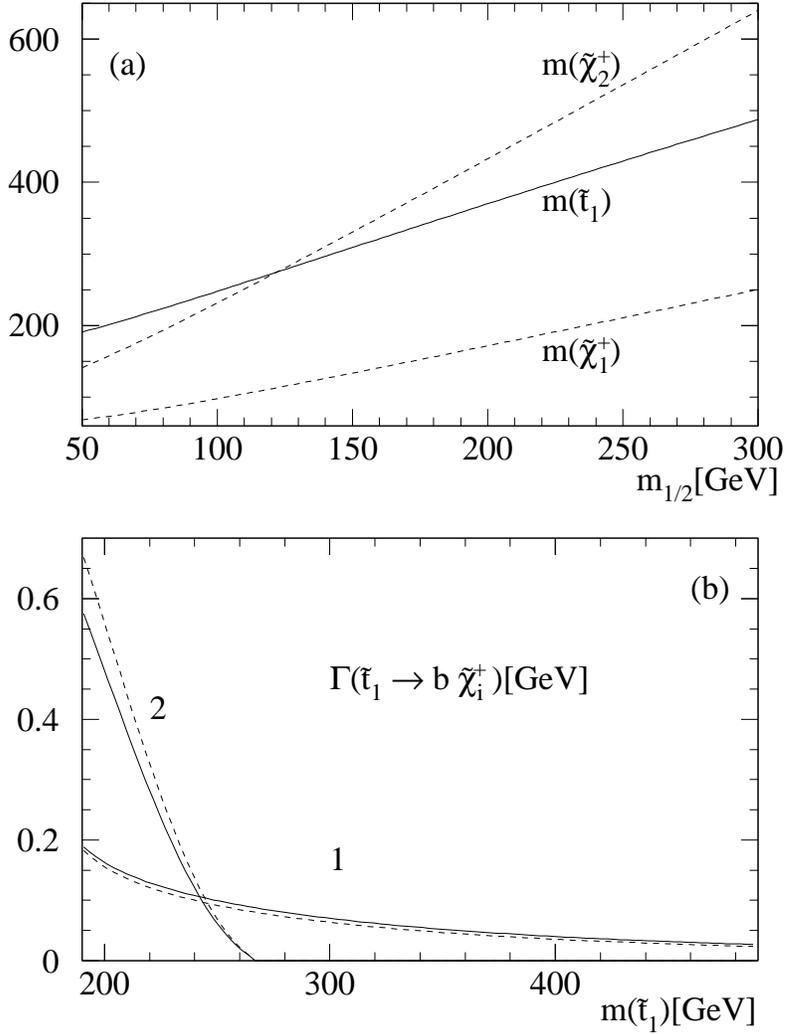,width=12cm} 
\caption[]{The SUSY-QCD corrections to the decay of light stop particles into
bottom quarks and chargino eigenstates [see also 
Ref.\cite{vienna}]:
(a) the masses of the particles [in GeV] as a function of the 
common gaugino mass $m_{1/2}$;
(b) the decay widths in leading order (dashed curve) and next-to-leading 
order (solid curve).
Input parameter set: $m_0=50$~GeV, $A_0=100$~GeV, $\mu<0$.}
\label{fig_sc}
\end{figure}

For the sake of comparison we have re-analyzed stop decays into
neutralinos and charginos. The results agree with the analysis in
Ref.\cite{vienna}, and also with the parallel calculation in
Ref.\cite{karlsruhe}, with which detailed point-by-point comparisons
have been performed.  For the set of parameters chosen in the figures,
only two neutralino decay modes are kinematically allowed.  The SUSY-QCD
corrections to the decays into neutralinos are small [Fig.\ref{fig_sn}(b)], 
typically less than $10 \%$.  
The picture is quite similar for the decays
into charginos [Fig.\ref{fig_sc}(b)], though the corrections are
slightly larger as a result of the massless $b$ quarks in the final
state.\bigskip

{\boldmath $\text{\bf (d)} \; t \rightarrow  \ste + \chijn \qquad 
                                                     [j=1,...] $}

\hspace{0.6cm} for $m_0=250$~GeV, $A_0=800$~GeV, $\mu>0$

\begin{figure}[h]
\epsfig{file=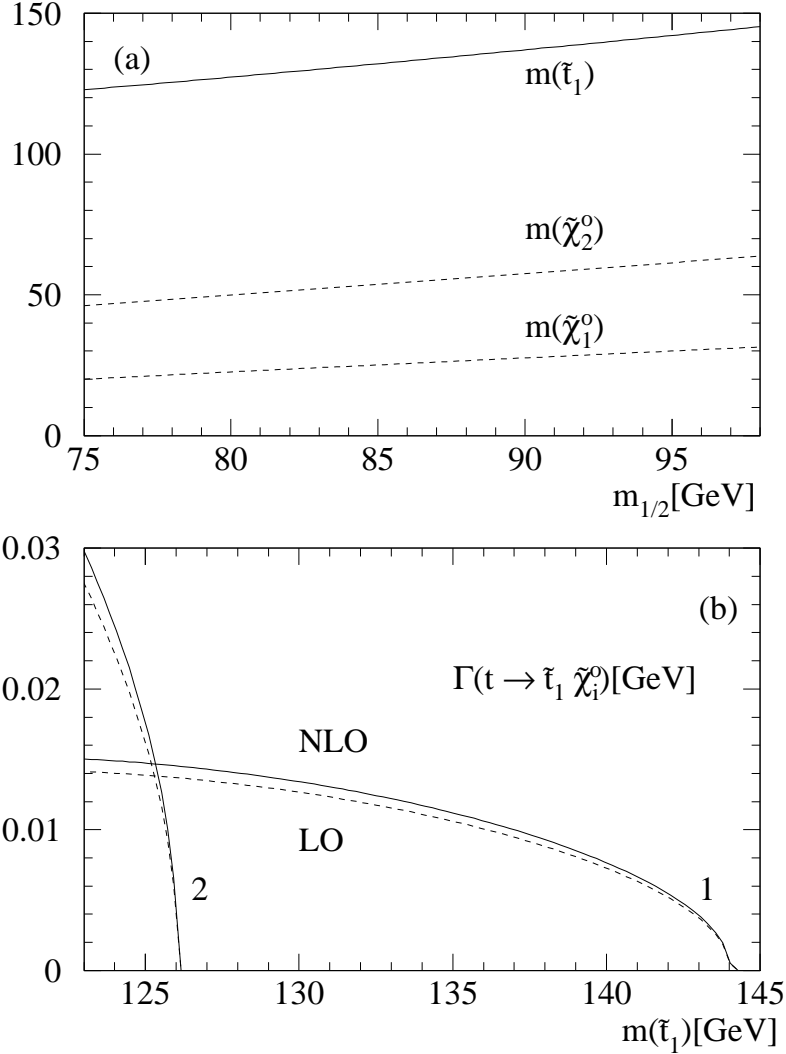,width=12cm} 
\caption[]{The SUSY-QCD corrections to the decay of top quarks into  
light $\ste$ particles and light neutralinos [see also 
Ref.\cite{karlsruhe}]:
(a) the masses of the particles [in GeV] as a function of the 
common gaugino mass $m_{1/2}$;
(b) the decay widths in leading order (dashed curve) and next-to-leading 
order (solid curve).
Input parameter set: $m_0=250$~GeV, $A_0=800$~GeV, $\mu>0$.
Only the decays into the two lightest neutralinos are kinematically allowed 
for this parameter set.}
\label{fig_tn}
\end{figure}

As in the previous case (c), this last example is shown merely as a
cross-check with Ref.\cite{top_decay}.  As shown in
Fig.\ref{fig_tn}(b), the corrections are small. If this top-decay mode
is realized in nature, the branching ratio for decays into $\chien$
can in principle be of the order of $4 \%$. 
[Note that $\Gamma(t \rightarrow bW^+)
\simeq 1.4$~GeV for $m_t = 175$~GeV \cite{bigi}].\bigskip

\setlength{\parindent}{5mm}

Since the $\stj/\gt/t$ SUSY decay modes could only be illustrated for
a specific set of parameters, a general program\footnote{The FORTRAN
  program may be obtained from plehn@@desy.de.} has been constructed for
generating all relevant decay widths in the stop--top sector.

\section{Summary}

In this paper we have analyzed the SUSY-QCD corrections for stop
decays to top quarks and gluinos, and gluino decays to stop particles
and top quarks. In contrast to earlier analyses, the non-zero top-quark 
mass must be taken into account in these decay modes. Moreover,
the $L/R$ squark mixing plays an important role.

We have set up a scheme in which the mixing angle $\tmix(Q^2)$ is
defined in such a way that the virtual/real stop particles $\ste$ and $\stz$
do not oscillate for a specific value of the four-momentum squared
$Q^2$.  Convenient choices for $Q^2$ are $Q^2=\mse^2$ or $\msz^2$,
depending on the problem treated in the analysis. Different
conventions are connected by simple relations between
the associated mixing angles.

As observed earlier for squarks related to the light quark species,
the SUSY-QCD corrections are large and positive for stop decays to top
quarks and gluinos. They are modest and negative for gluino decays to
stop particles and top quarks. We have compared these modes with other
stop-decay modes, which have been analyzed earlier
\cite{vienna,karlsruhe}.  In contrast to the strong-interaction decays,
the SUSY-QCD corrections for electroweak stop decays into neutralinos and
charginos are small, and so are the corrections for top-quark decays
to stop particles and the lightest neutralino.

\section{Acknowledgments}
We are grateful to the authors of Ref.\cite{karlsruhe}, especially
A.~Djouadi and C.~J\"unger, for the mutual comparison of results on
the electroweak stop decays, and for general discussions on the
problems treated in this paper.  We thank P.~Ohmann for providing us
with a program in which the renormalization-group equations are solved
for the masses of supersymmetric particles in supergravity scenarios.
Special thanks go also to A.~Bartl, A.~Djouadi and W.~Majerotto for
valuable comments on the manuscript.

\section{Appendix}
\small

In this appendix we present the explicit formulae for the calculation
of the decay width of $\ste$ particles to top quarks and gluinos in
next-to-leading order SUSY-QCD. The results for the decay of $\stz$
particles can be derived by interchanging the stop masses and
switching the sign in front of $\sin(2 \tmix)$ and $\cos(2 \tmix)$.
The gluino decay width $\Gamma (\gt \rightarrow \tb\,\stj)$ can be
obtained from $\Gamma (\stj \rightarrow t\gt )$ by adding a factor
$-\msj^3/(4C_F \mg^3)$ and subsequent analytical continuation from the
region $\,\msj > \mt + \mg\,$ to the region $\,\mg > \mt + \msj$.  The
multiplicative factor reflects the difference in phase-space, in
color/spin averaging, and in the sign of the gluino momentum inside
the spinor
sum [see also Eqs.~(\ref{eq_gamlo_stj}) and (\ref{eq_gamlo_gt})]. 
\bigskip

The decay width in next-to-leading order may be split into the
following components:
\begin{equation*}
\Gamma_{\text{NLO}} = \Gamma_{\text{LO}} 
                    + \real [\Delta \Gamma_t 
                    + \Delta \Gamma_{\tilde{g}}
                    + \Delta \Gamma_{11}
                    + \Delta \Gamma_v 
                    + \Delta \Gamma_r
                    + \Delta \Gamma_c
                    + \Delta \Gamma_f                    
                    + \Delta \Gamma_{\text{dec}} ]                    
\end{equation*}

\setlength{\parindent}{0pt} To allow for more compact expressions we
first define a few short-hand notations:
\begin{alignat}{2}
&\mu_{abc} = m_a^2 + m_b^2 - m_c^2  \qquad
 \sigzt = \mt \mg \sin(2 \tmix) \nt[2mm]
&\fac = \frac{\kappa}{16 \pi \mse^3 N_c} \nt 
&\kappa = \left[ \mse^4 + \mg^4 + \mt^4 
               - 2 ( \mse^2 \mg^2 + \mse^2 \mt^2 + \mg^2 \mt^2 ) \right]^{1/2} 
\phantom{....................} \notag
\end{alignat}
where $abc = \gt,t,j$ with $j$ representing $\stj$.

We list the components defined above for the next-to-leading order 
decay width [the $1/\varepsilon$ poles in the scalar 
integrals cancel against each other in the final sum]:\smallskip

\underline{lowest-order decay width}:
\begin{equation*}
\Gamma_{\text{LO}} = 8 N_c C_F \pi\as \left( - \mugte + 2 \sigzt \right) \,\fac
                   \equiv \fac \MB
\end{equation*}

\underline{top self-energy contribution}: 
\begin{alignat}{4}
   \Delta \Gamma_t &=\:&
        \frac{2 C_F \fac}{\mt^2}\, \pi \as \MB  &\Big\{
            2 (1-\ee) A(\mt) 
          + 2 A(\mg)             
          - A(\mse)
          - A(\msz)                 \nt &&&
          + \muteg B(\pt^2,\mg,\mse) 
          + \mutzg B(\pt^2,\mg,\msz)          \nt[1mm] &&&
          - 4 \mt^2 \sigzt \big[ \Bp(\pt^2,\mg,\mse)  
                           - \Bp(\pt^2,\mg,\msz) \big]  \nt[1mm] &&&
          + 2 \mt^2 \big[ \mugte \Bp(\pt^2,\mg,\mse) 
                    + \mugtz \Bp(\pt^2,\mg,\msz) 
                    - 4 \mt^2 \Bp(\pt^2,\ml,\mt) \big]  
          \Big\} \nt 
       &\ +\:&  \frac{16 N_c C_F^2 \fac}{\mt^2}\, \pi^2 \as^2 \, \czt^2 &\,  
            \mugte \, \Big\{
            A(\msz) 
          - A(\mse)   
          - \mugte B(\pt^2,\mg,\mse) 
          + \mugtz B(\pt^2,\mg,\msz) 
          \Big\} \phantom{.........} \notag
\end{alignat}

\underline{gluino self-energy contribution} (for $n_f=6$ quark flavors):
\begin{alignat}{4}
   \Delta \Gamma_{\gt} &=\:&
         \frac{4 \fac}{\mg^2}\, \pi \as \MB &(n_f-1) \Big\{
          - A(\ms)
          + ( \ms^2
            + \mg^2 ) B(\pg^2,\ms,0)  
          + 2 \mg^2 ( \mg^2 - \ms^2 ) \Bp(\pg^2,\ms,0) 
          \Big\} \nt 
      &\ +\:& \frac{2 \fac}{\mg^2}\, \pi \as \MB &\Big\{
            2 A(\mt)
          - A(\mse)
          - A(\msz) 
          + \muegt B(\pg^2,\mse,\mt) 
          + \muzgt B(\pg^2,\msz,\mt) \nt &&& \phantom{\big[}
          - 4 \mg^2 \sigzt \big[ \Bp(\pg^2,\mse,\mt)
                    - \Bp(\pg^2,\msz,\mt) \big] \nt[1mm] &&& \phantom{\big[}
          + 2 \mg^2 \big[ \mugte \Bp(\pg^2,\mse,\mt)  
                        + \mugtz \Bp(\pg^2,\msz,\mt) \big]
           \Big\} \nt 
      &\ +\:& \frac{4 \fac}{\mg^2}\, \pi \as \MB  & N_c \, \Big\{
            (1-\ee) A(\mg)
          - 4 \mg^4 \Bp(\pg^2,\ml,\mg) \Big\}
       \phantom{...................} \notag
\end{alignat}

\underline{diagonal stop self-energy}:
\begin{alignat}{4}
   \Delta \Gamma_{11} &=\:&
         8 C_F \fac \pi \as \MB \Big\{&
            B(\ps^2,\mg,\mt)
          - B(\ps^2,\ml,\mse) 
          + 2 \sigzt \Bp(\ps^2,\mg,\mt) \nt[1mm] &&&
          - \mugte \Bp(\ps^2,\mg,\mt) 
          - 2 \mse^2 \Bp(\ps^2,\ml,\mse) 
          \Big\}  \nt[1mm] \notag 
\end{alignat}
[The off-diagonal mixing contribution 
\begin{alignat}{4}
   \Delta \Gamma_{12} &=\:&
        \frac{128 N_c C_F^2 \fac}{\mse^2-\msz^2}\, \pi^2 \as^2 \, \sigzt \, &
          \czt^2 \, \Big\{
          A(\msz) 
        - A(\mse) 
        + \frac{4 \mt^2 \mg^2}{\sigzt}\, B(\ps^2,\mg,\mt) 
        \Big\} \phantom{........} \notag
\end{alignat}
is absorbed into the renormalization of the mixing angle for 
$\tmix = \tmix(\mse^2)$.]

\underline{vertex corrections}:
\begin{alignat}{4}
   \Delta \Gamma_v &=\:&
      64 \fac \pi^2 \as^2 N_c C_F^2 & 
       \left[ F^F_1 
            + \sigzt F^F_2 
            + \sigzt^2 F^F_3 \right] \phantom{.............................} 
  \nt 
  &\ +\:& 32 \fac \pi^2 \as^2 N_c^2 C_F & 
       \left[ F^A_1 
            + \sigzt F^A_2 
            + \sigzt^2 F^A_3 \right] \nt 
  &\ +\:&  8 \fac \pi \as \MB & F^B \notag 
\end{alignat}
with:
\begin{alignat}{2}
   F^F_1 =\:&
            \:2 ( \mt^2 
              + \mg^2 ) B(\ps^2,\mg,\mt) 
          + ( \mse^2 
            + \mt^2
            + \mg^2 ) B(\ps^2,\ml,\mse) \nt &
          + 2 ( \mg^2 
              - \mse^2 ) B(\pt^2,\ml,\mt) 
          - 2 \mt^2 B(\pt^2,\mg,\msz) 
          - 4 \mg^2 B(\pg^2,\mt,\mse)  \nt &
          + 4 \mg^2 ( \mse^2 
                    - \mg^2 ) C(\ps,-\pt,\mt,\mg,\mse) 
          + 2 \mt^2 ( \mse^2 + \msz^2 - 2 \mt^2 ) C(\ps,-\pt,\mt,\mg,\msz) 
          \nt 
   F^F_2 =\:&
          - 2 B(\ps^2,\ml,\mse)
          - 2 B(\pt^2,\ml,\mt) 
          - 4 B(\ps^2,\mg,\mt)
          + 2 B(\pt^2,\mg,\mse) \nt & 
          + 4 B(\pg^2,\mt,\mse) 
          + 4 \mugte C(\ps,-\pt,\mt,\mg,\mse) 
          + 2 (\mse^2-\msz^2) C(\ps,-\pt,\mt,\mg,\msz)
          \nt
   F^F_3 =\:&
       \frac{1}{\mg^2 \mt^2} \Big\{
            2 \mt^2 [ B(\pt^2,\mg,\msz) 
                    - B(\pt^2,\mg,\mse) ] 
          + \mugte ( \mugte 
                   - 4 \mt^2 ) C(\ps,-\pt,\mt,\mg,\mse) \nt & \qquad \quad
          - ( \mugte \mugtz 
            - 2 \mt^2 \mugte
            - 2 \mt^2 \mugtz ) C(\ps,-\pt,\mt,\mg,\msz)  
          \Big\} \nt
   F^A_1 =\:&
          - 2 \ee \,\mugte B(\ps^2,\mg,\mt) 
          + 4 ( \mt^2 
              - \mse^2 ) B(\pg^2,\ml,\mg) \nt &
          + 2\mt^2 [ B(\pt^2,\mg,\msz) 
                  - B(\pt^2,\mg,\mse) ] 
          + 4 \mg^2 B(\pg^2,\mt,\mse) \nt &
          + 4 \mg^2 ( \mg^2 
                    - \mse^2 ) C(\ps,-\pt,\mt,\mg,\mse) 
          - 2 \mt^2 ( \mse^2 + \msz^2 - 2 \mt^2 ) C(\ps,-\pt,\mt,\mg,\msz)
           \nt 
   F^A_2 =\:&
            4 \ee \,B(\ps^2,\mg,\mt) 
          - 4 B(\pg^2,\ml,\mg)
          - 4 B(\pg^2,\mt,\mse) \nt &
          - 4 \mugte C(\ps,-\pt,\mt,\mg,\mse)
          - 2 (\mse^2-\msz^2) C(\ps,-\pt,\mt,\mg,\msz) 
          \nt 
   F^A_3 =\:&
       \frac{1}{\mg^2 \mt^2} \Big\{
            2 \mt^2 [ B(\pt^2,\mg,\mse) 
                    - B(\pt^2,\mg,\msz) ] 
          + \mugte ( 4 \mt^2 
                   - \mugte ) C(\ps,-\pt,\mt,\mg,\mse) \nt & \qquad \quad
          + ( \mugte \mugtz 
            - 2 \mt^2 \mugte
            - 2 \mt^2 \mugtz ) C(\ps,-\pt,\mt,\mg,\msz) 
          \Big\} \nt 
   F^B =\:&
          N_c \big[
            \muteg C(\ps,-\pt,\mse,\ml,\mt) 
          - \mugte C(\ps,-\pt,\mg,\mt,\ml) 
          - \muegt C(\ps,-\pt,\ml,\mse,\mg) \big] \nt &  
          - 2 C_F \muteg C(\ps,-\pt,\mse,\ml,\mt) \notag
\end{alignat}

\underline{corrections from real-gluon radiation}:
\begin{alignat}{2} 
\Delta \Gamma_r &= 
         \frac{\as}{4 \pi^2 \mse} \MB \left[ 
                                       ( \mse^2   
                                       - \mt^2 ) I_{\ste \gt}  
                                       - \mt^2 I_{\ste t}  
                                       - \mg^2 I_{\gt \gt}  
                                       - I_{\gt}   \right] \nt &
       + \frac{\as C_F}{4 \pi^2 \mse N_c} \MB \left[  
                                       {}- \mse^2 I_{\ste \ste}  
                                       - \mt^2 I_{t t}  
                                       + \muteg I_{\ste t}  
                                       + I_{\ste}  
                                       - I_t \right] 
       \phantom{....................} \nt &
       + \frac{\as^2 C_F}{\pi \mse} \left[ C_F I^{\gt}_t  
                                         + N_c I^{\ste}_{\gt} \right] \notag 
\end{alignat}

\underline{renormalization of the coupling constant}:
\begin{equation*}
\Delta \Gamma_c = {}-\frac{\fac \as}{4 \pi} \MB
                   \left[ \frac{1}{\ee} 
                        - \gamma_E 
                        + \log(4 \pi) 
                        - \log\left(\frac{\mu_R^2}{\mu^2}\right) \right] 
                                          \left( \frac{11}{3}N_c 
                                               - \frac{2}{3}N_c 
                                               - \frac{2}{3}n_f 
                                               - \frac{1}{3}n_f \right) 
\end{equation*}

\underline{finite shift of the Yukawa coupling relative to the gauge
  coupling in $\msbar$}:
\begin{equation*}
\Delta \Gamma_f = \frac{\fac \as}{4 \pi} \MB 
                 \left( \frac{4}{3} N_c - C_F \right)
\end{equation*}

\underline{decoupling of the heavy flavors from the running strong coupling}:  
\begin{alignat}{4}
\Delta \Gamma_{\text{dec}} &=\:& \frac{\fac \as}{\pi} \MB \left\{ 
      \vphantom{\log\left(\frac{\mu_R^2}{\ms^2}\right)} \right.&
      \frac{n_f-1}{12} \log\left(\frac{\mu_R^2}{\ms^2}\right) 
    + \frac{1}{24} \log\left(\frac{\mu_R^2}{\mse^2}\right)   
    + \frac{1}{24} \log\left(\frac{\mu_R^2}{\msz^2}\right) 
    \phantom{ .................... } \nt &&& \left. \! 
    {}+ \frac{1}{6} \log\left(\frac{\mu_R^2}{\mt^2}\right)    
    + \frac{N_c}{6} \log\left(\frac{\mu_R^2}{\mg^2}\right) 
    \right\} \notag 
\end{alignat}

In these expressions the invariant integrals are defined as
\begin{alignat}{2}
&I^{{_A}{_B}...}_{ab...} = \frac{1}{\pi^2}\,\int \frac{d^3 \pt}{2 p_t^0} 
                           \frac{d^3p_{\gt}}{2p_{\gt}^0}
                           \frac{d^3q}{2q^0}
                     \delta_4 \left( q+\pt+\pg-\ps \right)
                     \frac{(2qp_{_A})(2qp_{_B}) \cdots}{(2qp_a)(2qp_b) \cdots}
                              \notag \\
&A(m_a) = \mu^{2\ee}\int \frac{d^nq}{(2\pi)^n} \frac{i}{q^2-m_a^2} \notag \\
&B(p^2,m_a,m_b) = \mu^{2\ee}\int \frac{d^nq}{(2\pi)^n} 
            \frac{i}{[q^2-m_a^2][(q+p)^2-m_b^2]} \notag \\
&C(p_1,p_2,m_a,m_b,m_c) = \mu^{2\ee}\int \frac{d^nq}{(2\pi)^n} 
        \frac{i}{[q^2-m_a^2][(q+p_1)^2-m_b^2][(q+p_1+p_2)^2-m_c^2]} 
                              \notag \\[1mm]
&\dot{B}(p^2,m_a,m_b) = \partial B(p^2,m_a,m_b)/\partial p^2 \notag
\end{alignat}
Note that in the angular integrals $I^{{_A}{_B}...}_{ab...}$ from real-gluon
radiation we do not add a minus sign for every term 
$(2qp_{\ste})$, in contrast to Ref.\cite{denner}.%
\footnote{It should be noted that Eqs.(D.11) and (D.12) of
  Ref.\cite{denner} must be corrected in the following way: The first
  term in the square brackets of Eq.(D.11) has to be replaced by the
  corresponding term in Eq.(D.12), and \emph{vice versa}.} In the
numerical analyses we have inserted the mass of the decaying
particle for the renormalization scale $\mu_R$. The scale parameter $\mu$
accounts for the correct dimension of the coupling in $n$ dimensions,
cancelled together with the $1/\varepsilon$ poles.  \normalsize

 \newcommand{\zp}[3]{{\sl Z. Phys.} {\bf #1} (19#2) #3}
 \newcommand{\np}[3]{{\sl Nucl. Phys.} {\bf #1} (19#2)~#3}
 \newcommand{\pl}[3]{{\sl Phys. Lett.} {\bf #1} (19#2) #3}
 \newcommand{\pr}[3]{{\sl Phys. Rev.} {\bf #1} (19#2) #3}
 \newcommand{\prl}[3]{{\sl Phys. Rev. Lett.} {\bf #1} (19#2) #3}
 \newcommand{\prep}[3]{{\sl Phys. Rep.} {\bf #1} (19#2) #3}
 \newcommand{\fp}[3]{{\sl Fortschr. Phys.} {\bf #1} (19#2) #3}
 \newcommand{\nc}[3]{{\sl Nuovo Cimento} {\bf #1} (19#2) #3}
 \newcommand{\ijmp}[3]{{\sl Int. J. Mod. Phys.} {\bf #1} (19#2) #3}
 \newcommand{\ptp}[3]{{\sl Prog. Theo. Phys.} {\bf #1} (19#2) #3}
 \newcommand{\sjnp}[3]{{\sl Sov. J. Nucl. Phys.} {\bf #1} (19#2) #3}
 \newcommand{\cpc}[3]{{\sl Comp. Phys. Commun.} {\bf #1} (19#2) #3}
 \newcommand{\mpl}[3]{{\sl Mod. Phys. Lett.} {\bf #1} (19#2) #3}
 \newcommand{\cmp}[3]{{\sl Commun. Math. Phys.} {\bf #1} (19#2) #3}
 \newcommand{\jmp}[3]{{\sl J. Math. Phys.} {\bf #1} (19#2) #3}
 \newcommand{\nim}[3]{{\sl Nucl. Instr. Meth.} {\bf #1} (19#2) #3}
 \newcommand{\el}[3]{{\sl Europhysics Letters} {\bf #1} (19#2) #3}
 \newcommand{\ap}[3]{{\sl Ann. of Phys.} {\bf #1} (19#2) #3}
 \newcommand{\jetp}[3]{{\sl JETP} {\bf #1} (19#2) #3}
 \newcommand{\jetpl}[3]{{\sl JETP Lett.} {\bf #1} (19#2) #3}
 \newcommand{\acpp}[3]{{\sl Acta Physica Polonica} {\bf #1} (19#2) #3}
 \newcommand{\vj}[4]{{\sl #1~}{\bf #2} (19#3) #4}
 \newcommand{\ej}[3]{{\bf #1} (19#2) #3}
 \newcommand{\vjs}[2]{{\sl #1~}{\bf #2}}
 \newcommand{\hep}[1]{{\sl hep--ph/}{#1}}
 \newcommand{\desy}[1]{{\sl DESY-Report~}{#1}}

\newpage

\end{document}